# Co-nondeterminism in compositions: A kernelization lower bound for a Ramsey-type problem[*]

Stefan Kratsch[†]

August 27, 2018

**Abstract**

Until recently, techniques for obtaining lower bounds for kernelization were one of the most sought after tools in the field of parameterized complexity. Now, after a strong influx of techniques, we are in the fortunate situation of having tools available that are even stronger than what has been required in their applications so far. Based on a result of Fortnow and Santhanam (STOC 2008, JCSS 2011), Bodlaender et al. (ICALP 2008, JCSS 2009) showed that, unless NP $\subseteq$ coNP/poly, the existence of a deterministic polynomial-time composition algorithm, i.e., an algorithm which outputs an instance of bounded parameter value which is yes if and only if one of t input instances is yes, rules out the existence of polynomial kernels for a problem. Dell and van Melkebeek (STOC 2010) continued this line of research and, amongst others, were able to rule out kernels of size $\mathcal{O}(k^{d-\epsilon})$ for certain problems, assuming NP $\nsubseteq$ coNP/poly. It is an immediate consequence of their work that even the existence of a co-nondeterministic composition rules out polynomial kernels. However, in contrast to the numerous applications of deterministic composition, the added power of co-nondeterminism has not yet been harnessed to obtain kernelization lower bounds.

In this work we present the first example of how co-nondeterminism can help to make a composition algorithm. We study the existence of polynomial kernels for a Ramsey-type problem: Given a graph $G$ and an integer $k$, the question is whether $G$ contains an independent set or a clique of size at least $k$. It was asked by Rod Downey whether this problem admits a polynomial kernelization, and such a result would greatly speed up the computation of Ramsey numbers. We provide a co-nondeterministic composition based on embedding t instances into a single host graph $H$. The crux is that the host graph $H$ needs to observe a bound of $\ell \in \mathcal{O}(\log t)$ on both its maximum independent set and maximum clique size, while also having a cover of its vertex set by independent sets and cliques all of size $\ell$; the co-nondeterministic composition is build around the search for such graphs. Thus we show that, unless NP $\subseteq$ coNP/poly (and the polynomial hierarchy collapses), the problem does not admit a kernelization with polynomial size guarantee.

## 1 Introduction

Parameterized complexity refines classical complexity by taking into account not only the size of a given input but also one or more additional parameters like solution size, or structural measures like various notions of width for graphs. The main positive result that one seeks to obtain, is to show that instances $(x, k)$ of a given NP-hard problem can be solved in time $\mathcal{O}(f(k) \cdot |x|^c)$ where $f$ is a computable function and $c$ is a constant independent of $k$; this is called *fixed-parameter tractability*. It entails $\mathcal{O}(|x|^c)$ algorithms for every bounded value of $k$. If the chosen parameter $k$ can be expected to be small in practice, then this is a strong improvement over a worst-case exponential time, e.g., $\mathcal{O}(\alpha^{|x|})$, algorithm that one would otherwise have to resort to (given our current knowledge of P vs. NP and hypotheses like the exponential time hypothesis, cf. [22]).

---

[*]Supported by the Netherlands Organization for Scientific Research (NWO), project "KERNELS: Combinatorial Analysis of Data Reduction".

[†]Utrecht University, Utrecht, the Netherlands

Kernelization takes the perspective that if the chosen parameter $k$ is small when compared to the size of a given instance $(x, k)$, then strong insights into the structure of the instance should be possible which allow to discard large parts of $x$ in polynomial time and leave an equivalent instance of size bounded by some function in $k$. Interestingly, by a folklore result, the problems with such a kernelization are exactly those in the class FPT of fixed-parameter tractable problems. This shows that kernelization is a robust definition of data reduction, which is not possible when taking into account only the input size (see also the discussion by Harnik and Naor [13] in a study of compression related to witness size). An important subclass of FPT is formed by those problems allowing kernelizations with size guarantee polynomial in $k$, capturing plenty of results with linear or quadratic size kernels, e.g., [21, 3, 10], but enjoying the good closure properties of polynomials.

A nice feature of kernelization is that since many parameters can be well approximated, it is not necessary to follow up with an exact or FPT algorithm or even to adopt the framework of parameterized complexity in the first place. Since only polynomial time is invested to get the kernelized instance, it is just as valid to run an approximation, randomized, or heuristic algorithm afterwards. In fact, reduction rules have had fair use in other areas already and, e.g., primal-dual approximation techniques are quite related to standard arguments in kernelization which start from a packing of forbidden structures (see, for example, Paul et al. [18]).

Until recently, techniques for obtaining lower bounds for kernelization were one of the most sought after tools in the field of parameterized complexity (see, e.g., a 2007 survey of Guo and Niedermeier [12]). This was especially true for the threshold of whether or not a problem would allow a polynomial kernel. Now, after a strong influx of techniques [2, 11, 5, 7, 4, 14], we are in the fortunate situation of having tools that are even stronger than what has been required in their applications so far.

Let us take a high level view of the main technique for excluding polynomial kernels. The central piece is that of a *composition algorithm* which takes as input $t$ instances $(x_1, k), \ldots, (x_t, k)$ and produces in polynomial time an instance $(y, k')$ which is **yes** if and only at least one $(x_i, k)$ is **yes**, and, crucially, with $k'$ polynomially bounded in $k$. When combined with a polynomial kernelization this gives a *distillation algorithm* for the underlying classical problem which given $\tilde{x}_1, \ldots, \tilde{x}_t$ computes in polynomial time an instance $\tilde{y}$ which is **yes** if at least one $\tilde{x}_i$ is **yes**, and whose size is polynomially bounded in the largest $\tilde{x}_i$. The intuition of this framework given by Bodlaender et al. [2] is that when $t$ exceeds the size of $\tilde{y}$ (which is independent of $t$) then there will be less than one bit of information per instance; they conjectured that NP-hard problems do not have distillation algorithms. Fortnow and Santhanam [11] proved the conjecture to be true under the assumption that NP $\nsubseteq$ coNP/poly (known to otherwise cause a collapse of the polynomial hierarchy [23]). This led to flurry of papers showing composition algorithms for various problems, e.g., [8, 9, 5, 16], and thus excluding polynomial kernelizations assuming NP $\nsubseteq$ coNP/poly.

By a generalization of the work of Fortnow and Santhanam [11] Dell and van Melkebeek [7] show that languages $L$ which have an oracle communication protocol for deciding instances $(x_1, \ldots, x_t)$ of $OR(L)$ with (communication) cost $\mathcal{O}(t \log t)$ are contained in coNP/poly; given $(x_1, \ldots, x_t)$, the $OR(L)$ problem asks whether at least one $x_i$ is contained in $L$. They conclude that NP-hard languages $L$ do not have such protocols unless NP $\subseteq$ coNP/poly. Combined with an intricate packing lemma, this led to their main result that satisfiability of $d$-CNF formulas does not allow nontrivial sparsification, i.e., instances with $n$ variables cannot be compressed to size $\mathcal{O}(n^{d-\epsilon})$. Amongst other things, they also obtain polynomial lower bounds for kernelization, e.g., non-existence of a $\mathcal{O}(k^{2-\epsilon})$ sized kernel for VERTEX COVER (all results assuming NP $\nsubseteq$ coNP/poly). Combining a polynomial kernelization and a composition algorithm naturally gives an oracle communication protocol [7].

An interesting new aspect in the lower bounds via oracle communication protocols (see Section 3 for a definition) is that the exclusion of protocols of cost $\mathcal{O}(t \log t)$ holds, explicitly, even when the first player (holding the input and communicating with an all-powerful oracle) is allowed to behave co-nondeterministically [7]. The fact that co-nondeterminism can be allowed is already implicit in the work of Fortnow and Santhanam [11], as observed by Chen and Müller (cf. [13]). The key observation seems to be that, essentially, a kernel and a composition are used as subroutines in a coNP-machine for accepting an NP-hard language. Hence, relaxing the subroutines to co-nondeterministic behavior as well does not harm the properties of the accepting machine. To our knowledge, the only result so far making use of co-nondeterminism is the lower bound of $\mathcal{O}(n^{d-\epsilon})$ on PCPs for $d$-SAT [7]. In particular, the implicit notion of co-nondeterministic composition is left largely unexplored, despite of the high interest in a set of problems that so far resisted a classification into admitting or non admitting a polynomial kernelization, e.g., DIRECTED FEEDBACK VERTEX SET and MULTIWAY CUT. Building on the work of Dell and van Melkebeek [7], recent work of Hermelin and Wu [14] defines a notion called *weak composition* which permits a larger dependence on the number $t$ of instances. They obtain concrete polynomial lower bounds in the style of Dell and van Melkebeek [7], i.e., for problems which admit *some* polynomial kernel. By definition, weak compositions allow co-nondeterminism, but the current results make no use of this option. Our co-nondeterministic composition excludes kernels of *any* polynomial size.

**The Ramsey problem.** Recently, Rod Downey posed the interesting question of whether the following combination of the well-known CLIQUE and INDEPENDENT SET problems admits a polynomial kernel [17]. We call it RAMSEY($k$) for brevity.

> RAMSEY($k$)
> **Input:** A graph $G$ and an integer $k$.
> **Parameter:** $k$.
> **Question:** Does $G$ contain an independent set or clique of size $k$?

Unlike CLIQUE and INDEPENDENT SET, the problem is FPT by a more general result of Khot and Raman [15] which uses Ramsey's Theorem: Let $R(k)$ denote the smallest integer $N$ such that each graph with $N$ vertices contains an independent set or a clique of size $k$; Ramsey showed these numbers to exist and to be computable [19]. If $G$ has more than $R(k)$ vertices, then the instance is **yes**. Otherwise, the number of possible solutions is bounded by $f(k) = (R(k))^k$; since $R(k)$ is computable this suffices to prove fixed-parameter tractability (see Section 2 for explicit upper and lower bounds on $R(k)$). However, it is open whether or not there is a polynomial kernelization for it. The question of small kernels for the RAMSEY($k$) problem is well-motivated: There are as of yet no efficient algorithms known for computing Ramsey numbers; a brute-force way is to check all non-isomorphic graphs on $N$ vertices for $k$-cliques or $k$-independent sets in order to determine whether $R(k) \leq N$. The known bounds for $R(k)$ imply that this requires $N$ to be of order $\mathcal{O}(\alpha^k)$, giving a runtime of $\mathcal{O}(\alpha^{k^2})$ per graph (trying all sets of $k$ vertices). A polynomial kernelization which guarantees reduction to $\mathcal{O}(k^c)$ vertices would yield runtime $\mathcal{O}((k^c)^k) = \mathcal{O}(\alpha^{k \log k})$ per graph.

**Our work.** Regarding polynomial kernelization for RAMSEY($k$) we demonstrate two things. We disprove the existence of polynomial kernels for RAMSEY($k$) unless NP $\subseteq$ coNP/poly. We thereby show for the first time how to exploit co-nondeterminism to construct a composition algorithm. It appears that the co-nondeterminism is necessary to realize our composition algorithm, since it involves detection of cliques and independent sets (see below).

**Techniques and related work.** Unlike for the problems CLIQUE and $k$-PATH [13, 2], the disjoint union of $t$ instances of RAMSEY($k$) does not work satisfactorily as a composition algorithm (and neither would a join of the instances) as it would contain independent sets of size $\Omega(t)$. The intricate Packing Lemma due to Dell and van Melkebeek [7, Lemma 1], designed of course for a different purpose, does not seem to be applicable either as it constructs an $n$-partite graph containing independent sets of size $n$ which cannot be bounded in $\mathcal{O}(\log t)$ when $t := t(n)$ is polynomially-bounded. Generally, it appears to be unlikely that one could pack the instances in such a way that solutions are confined to a part representing a single original instance.

Our construction can best be motivated by a simplified example. Let $t = \ell^2$ instances of RAMSEY($k$) be given, say, $(G_1, k), \ldots, (G_t, k)$, and assume that each instance contains at least one independent set and one clique of size $k - 1$. We construct a graph $G'$ as follows: Let $G'$ contain copies of the graphs $G_1, \ldots, G_t$, and pick an arbitrary partition of the graphs into $\ell$ groups of size $\ell$ each. Then add all edges between vertices of different graphs that are in the same group. Now, if all $t$ instances are **no**, then it can be verified that $G'$ contains no clique and no independent set of size greater than $\ell \cdot (k - 1)$: The reason is that any clique or independent set can contain vertices from at most $\ell$ graphs $G_i$ (each clique only from one group; each independent set only from one graph per group). If at least one instance is **yes** then its independent set or clique of size $k$ can be extended with $k - 1$ vertices of each of $\ell - 1$ other graphs; this gives a solution of size $\ell \cdot (k - 1) + 1$. Thus asking whether $G'$ has an independent set or clique of size at least $\ell \cdot (k - 1) + 1 \in \mathcal{O}(\sqrt{t}k)$ is equivalent to whether at least one instance $(G_i, k)$ is **yes**. We mention in passing that such a composition excludes kernels of size $\mathcal{O}(k^{2-\epsilon})$ by recent work of Hermelin and Wu [14], or by deriving an appropriate communication protocol and applying the mentioned result of Dell and van Melkebeek [7].

The reader may have noticed that in the example we have connected the instances according to the complement of the Turán graph $T(t, \ell)$ which (for $t = \ell^2$) contains no independent set or clique of size greater than $\ell$. The other equally important feature of the Turán graph that we exploited is that each vertex is contained in both an independent set and a clique of size exactly $\ell$. This way the distinction whether or not any one graph $G_i$ has a solution of size $k$ (instead of just $k - 1$) makes the crucial difference for the instance $(G', \ell(k - 1) + 1)$. Motivated by this example the main work lies in finding a better *host graph* $H$ to replace $T(n, \ell)$ which has similar properties but with $\ell \in \mathcal{O}(\log t)$. No deterministic construction is known for such graphs, despite fairly recent progress on deterministic construction of Ramsey graphs without cliques or independent sets of size $t^* + 1 = t^{o(1)}$ [1]. While $\ell = t^*$ can be seen to still exclude polynomial kernels (cf. Section 3), it seems unlikely that those graphs would support a cover with cliques and independent sets each of size $t^*$; also, our tighter logarithmic dependence on $t$ may have other consequences for kernels. We ensure the covering property by using gaps between Ramsey numbers $R(\ell)$ and $R(\ell+1)$ when $\ell \in \mathcal{O}(\log t)$. This in turn would require deterministic constructions for $\mathcal{O}(\log t)$-Ramsey graphs which is open.

**Organization.** In Section 2 we recall the necessary definitions, mention upper and lower bounds on Ramsey numbers, and introduce a refinement version of RAMSEY($k$) which will be used for the composition. In Section 3, we state the required result of Dell and van Melkebeek [7], introduce the notion of co-nondeterministic composition which we will use, and show that this concept excludes polynomial kernels, assuming NP $\not\subseteq$ coNP/poly. In Section 4 we show an embedding of graphs into a *host graph*, motivated by the example using the edge complement of a Turán graph, but somewhat tweaked to lessen the restriction on the host graph. Section 5 then gives the co-nondeterministic composition and derives our main result. We conclude in Section 6.

## 2 Preliminaries

**Graphs.** All graphs considered in this work are finite, simple, and undirected. By the *join* of two graphs (or two connected components), we mean the operation of adding all edges between vertices of different graphs (or components). With $\alpha(G)$ and $\omega(G)$ we denote the maximum size of independent sets or cliques in $G$, respectively.

**Ramsey numbers.** The *Ramsey number* $R(k)$ is the smallest integer such that every graph on $R(k)$ vertices contains a clique or an independent set of size $k$. Ramsey's Theorem [19] shows that this number is finite. Currently the best bounds on these diagonal Ramsey numbers are as follows: Providing an upper bound, Conlon [6] shows that there is a constant $D$, such that for sufficiently large $k \in \mathbb{N}$ we have

$$R(k+1) \leq k^{-D\frac{\log k}{\log \log k}} \binom{2k}{k}.$$

Spencer [20] shows with an application of Lovász' Local Lemma that

$$R(k) > k2^{k/2}\left(\frac{1}{e\sqrt{2}} + o(1)\right).$$

**Parameterized problems and kernels.** A *parameterized problem* $\mathcal{Q}$ over alphabet $\Sigma$ is a subset of $\Sigma^* \times \mathbb{N}$. The problem $\mathcal{Q}$ is *fixed-parameter tractable* if there exists an algorithm $A$, a computable function $f$, and a constant $c$, such that $A$ decides membership in $\mathcal{Q}$ for any instance $(x, k)$ in time $\mathcal{O}(f(k)n^c)$. The problem $\mathcal{Q}$ admits a *kernelization* (or *kernel*) if there is a polynomial-time algorithm $K$ and a computable function $h$, such that $K$ transforms any instance $(x, k)$ into an equivalent instance $(x', k')$ with $|x'|, k' \leq h(k)$. The function $h$ is called the *size* of the kernelization $K$ and we say $K$ is a *polynomial kernelization* if $h(k)$ is polynomially bounded.

**Refinement version of Ramsey($k$).** Instead of considering RAMSEY($k$) directly, we focus on the following refinement version, in which the given graph is guaranteed to contain both a clique and an independent set of size $k-1$ (for ease of notation we omit the details of giving the $k-1$-sized independent set and clique in the input). Bodlaender et al. [2] use such problem variants to exclude, e.g., polynomial kernels for INDEPENDENT SET PARAMETERIZED BY TREEWIDTH.

REFINEMENT RAMSEY($k$)
**Input:** A graph $G$ and an integer $k$, such that $G$ has both an independent set and a clique of size $k - 1$.
**Parameter:** $k$.
**Question:** Does $G$ contain an independent set or clique of size $k$?

A simple reduction from RAMSEY($k$) to REFINEMENT RAMSEY($k$) which only increases the parameter by one shows that lower bounds transfer directly from the latter to the former problem; it is useful to note that instances for REFINEMENT RAMSEY($k$) are also legal for RAMSEY($k$), and applying the latter gives the same answer. We will use this later to transfer our obtained lower bound from REFINEMENT RAMSEY($k$) to RAMSEY($k$) (a more general argument for transferring lower bounds is due to Bodlaender et al. [5]).

**Lemma 1.** *There is a polynomial-time reduction reducing any instance $(G, k)$ of* RAMSEY($k$) *to an equivalent instance $(G', k + 1)$ of* REFINEMENT RAMSEY($k$).

*Proof.* Given an instance $(G, k)$ of RAMSEY($k$), and assuming w.l.o.g. that $k \geq 3$, construct $G'$ starting with a copy of $G$. Add a clique $C$ on $k - 1$ vertices to $G'$. Then add an independent set $I$ with $k$ vertices to $G'$ and make a join with all other vertices of $G'$ (in the copy of $G$ and in the clique $C$). Return $(G', k + 1)$.

If $G$ contains a $k$-clique, then in $G'$ a vertex of $I$ can be added to this clique to obtain a $k+1$-clique; if it contains a $k$-independent set then in $G'$ a vertex of $C$ can be added. Conversely, if $G'$ has a $k + 1$-clique $C'$, then this clique contains at most one vertex of $I$. Furthermore $C'$ cannot intersect $C$, else it could contain no vertex of the copy of $G$ limiting its size to $k$ (including the one vertex of $I$); thus $C'$ contains a $k$-clique in the copy of $G$. Similarly, if $G'$ contains a $k + 1$-independent set $I'$ then it cannot contain vertices of $I$, otherwise it could contain no further vertices due to the join operation. Thus it contains at most one vertex of the clique $C$ and an independent set of size at least $k$ in the copy of $G$. Finally, we observe that $G'$ contains a $k$-independent set, namely $I$, and a $k$-clique, formed by $C$ plus an arbitrary vertex of $I$. This completes the proof. □

We give a straightforward proof for NP-hardness of REFINEMENT RAMSEY($k$) and RAMSEY($k$). This is a prerequisite for the lower bound tools.

**Theorem 1.** *The problems* RAMSEY($k$) *and* REFINEMENT RAMSEY($k$) *are hard for* NP.

*Proof.* We give a reduction from CLIQUE. Let $(G, k)$ be an instance of CLIQUE, where $G$ has $n$ vertices. We construct a graph $G'$ by adding to $G$ a clique $C$ on $n + 1$ vertices, and adding all edges between the vertices of $G$ and $C$ (i.e., we perform a join operation on $G$ and $C$). We return $(G', k + n + 1)$ and claim that it is an equivalent instance of RAMSEY($k$).

Clearly the maximum clique size $\omega(G')$ of $G'$ is equal to $\omega(G) + n + 1$. We note also that the maximum independent size $\alpha(G')$ of $G'$ is at most $n$, since independent sets in $G'$ can either use the vertices of $G$ or a single vertex of the clique $C$.

Thus if $(G, k)$ is a **yes**-instance then $\omega(G) \geq k$ and $\omega(G') \geq k + n + 1$, and $(G', k + n + 1)$ is a **yes**-instance too. On the other hand, if $(G', k + n + 1)$ is a **yes**-instance then $\omega(G') \geq k + n + 1$ since $\alpha(G') \leq n$, implying that $\omega(G) \geq k$ and that $(G, k)$ is a **yes**-instance. Thus RAMSEY($k$) is NP-hard. NP-hardness of REFINEMENT RAMSEY($k$) now follows from Lemma 1. □

# 3  Lower bounds for kernelization

In this section we briefly recall the relevant results and definitions required to obtain our lower bound result. The main tool is the following lemma due to Dell and van Melkebeek [7]. Before stating the lemma, we recall their definition of an *oracle communication protocol*.

**Definition 1** ([7])**.** *An* oracle communication protocol *for a language L is a communication protocol for two players. The first player is given the input x and has to run in time polynomial in the length of the input; the second player is computationally unbounded but is not given any part of x. At the end of the protocol the first player should be able to decide whether $x \in L$. The cost of the protocol is the number of bits of communication from the first player to the second player.*

**Lemma 2** ([7])**.** *Let L be a language and $t\colon \mathbb{N} \to \mathbb{N} \setminus \{0\}$ be polynomially bounded such that the problem of deciding whether at least one out of $t(s)$ inputs of length at most s belongs to L has an oracle communication protocol of cost $\mathcal{O}(t(s)\log t(s))$, where the first player can be co-nondeterministic. Then $L \in \text{coNP}/\text{poly}$.*

It is an easy consequence of Lemma 2 that co-nondeterministic compositions lead to kernelization lower bounds. Being one of many other applications this extension is not made explicit by Dell and van Melkebeek [7] (though deterministic compositions are discussed), but their work motivated our search for a co-nondeterministic composition. Somewhat surprisingly, from sketching a proof for self-containment, it turns out that Lemma 2 not only permits co-nondeterminism. In fact, compositions with a dependence of $t^{o(1)}$ on $t$ can be showed to still exclude polynomial kernels (in [4] only a factor of $\log^c t$ is permitted for cross-compositions, and it comes from a different argument). Hermelin and Wu [14] gave a similar (if less explicit on coNP) proof for their notion of *weak composition* where $k' = t^{1/d}k^{\mathcal{O}(1)}$, showing that it excludes kernels of size $\mathcal{O}(k^{d-\epsilon})$. Their proof also allows $k' = t^{1/d+o(1)}k^{\mathcal{O}(1)}$.

We first give a definition of the version of composition that we are going to use.

**Definition 2.** *Let $\mathcal{Q} \subseteq \Sigma^* \times \mathbb{N}$. A co-nondeterministic polynomial-time algorithm $C$ is a* coNP-composition *for $\mathcal{Q}$ if there is a polynomial p such that on input of t instances $(x_1, k), \ldots, (x_t, k) \in \Sigma^* \times \mathbb{N}$ the algorithm $C$ takes time polynomial in $\sum_{i=1}^{t} |x_i|$ and outputs on each computation path an instance $(y, k') \subseteq \Sigma^* \times \mathbb{N}$ with $k' \leq t^{o(1)}p(k)$ and such that the following holds:*

- *If at least one instance $(x_i, k)$ is a **yes**-instance then all computation paths lead to the output of a **yes**-instance $(y, k')$.*

- *Otherwise, if all instances $(x_i, k)$ are **no**-instances, then at least one computation path leads to the output of a **no**-instance.*

We require the following notion of Bodlaender et al. [2] to state our lemma: The *unparameterized version* $\tilde{\mathcal{Q}}$ of a parameterized problem $\mathcal{Q}$ is defined as $\tilde{\mathcal{Q}} := \{x \# 1^k \mid (x, k) \in \mathcal{Q}\}$. It is essentially the same as $\mathcal{Q}$ except for the unary encoding of the parameter value, affecting its classical complexity.

**Lemma 3.** *Let $\mathcal{Q} \subseteq \Sigma^* \times \mathbb{N}$ be a parameterized problem such that $\tilde{\mathcal{Q}}$ is NP-hard. If $\mathcal{Q}$ has a coNP-composition then it does not admit a polynomial kernelization unless $\text{NP} \subseteq \text{coNP}/\text{poly}$ and the polynomial hierarchy collapses to its third level.*

*Proof.* Assume that $\mathcal{Q}$ admits a polynomial kernelization $K$ with polynomially bounded size $h$, say $h(k) = \mathcal{O}(k^c)$. Furthermore, let $C$ be a coNP-composition for $\mathcal{Q}$ which outputs instances with parameter bounded by $t^{o(1)}k^d$. We define a polynomially bounded function $t$ by $t(N) := N^{cd+2}$. By Lemma 2 it suffices to provide an oracle communication protocol for $\tilde{\mathcal{Q}}$ where the first player is co-nondeterministic and with cost $\mathcal{O}(t(N)\log t(N))$ for $t$ inputs each of size at most $N$.

Fixing $N$ and $t := t(N)$, let $t$ instances each of size at most $N$ be given to the first player, say $\tilde{x}_1, \ldots, \tilde{x}_t$. Let $(x_1, k_1), \ldots, (x_t, k_t)$ denote the corresponding parameterized instances of $\mathcal{Q}$.

Let us go through the protocol, but consider only the communication cost (for now). By definition of $\tilde{\mathcal{Q}}$ it follows that all $k_i$ are bounded by $N$. The first player groups the instances by parameter value (at most $N$ groups), and applies the co-nondeterministic composition to each group. In each computation path this gives $r \leq N$ instances $(G'_1, k'_1), \ldots, (G'_r, k'_r)$. Let us bound the parameter values $k'_i$, assuming that $(G'_i, k'_i)$ was obtained by composing all instances with parameter value $\hat{k}$:

$$k'_i = t^{o(1)}\hat{k}^d \leq t^{o(1)}N^d.$$

Now the first player applies the assumed polynomial kernelization to each instance $(G'_i, k'_i)$. Then he sends the obtained kernels to the second player, who tests membership for $\mathcal{Q}$ for each one. The second player answers **yes** if at least one of the instances send to him is **yes**, and **no** otherwise.

Each kernelized instance has size at most $h(k'_i) = \mathcal{O}((k'_i)^c) = \mathcal{O}((t^{o(1)}N^d)^c)$. Thus we can bound the cost of sending the at most $N$ kernelized instances to the second player as follows:

$$\mathcal{O}(N(t^{o(1)}N^d)^c) = \mathcal{O}(N^{cd+1}(t^{o(1)})^c) = \mathcal{O}(t),$$

using that $t = N^{cd+2}$.

It remains to show correctness, in particular taking into account the co-nondeterministic behavior of the composition. If at least one input instance $\tilde{x}_i$ is a **yes**-instance, then the corresponding instance $(G'_j, k'_j)$ will be **yes** on each computation path. Thus the oracle will answer **yes** on each computation path. Otherwise, if all instances are **no**, then there must be at least one computation path in which all $N$ runs of the coNP-composition return **no**-instances. Applying the kernelization will thus create $N$ **no**-instances as well (but note that a coNP-kernelization would suffice). These are then send to the oracle, causing it to answer **no** (on at least one path).

Thus, assuming a polynomial kernelization for $\mathcal{Q}$, we get an oracle communication protocol for deciding the OR of $t$ instances of $\tilde{\mathcal{Q}}$ of cost $\mathcal{O}(t)$. By Lemma 2 this implies that $\tilde{\mathcal{Q}}$ is contained in coNP/poly, and hence, by NP-hardness of $\tilde{\mathcal{Q}}$, that NP $\subseteq$ coNP/poly. □

## 4 The embedding construction

In this section we will describe the embedding to be used in the composition algorithm once a suitable *host graph* is found. Given $t$ instances of RAMSEY($k$), the construction requires a host graph $H$ with at least $t$ vertices. Furthermore, an integer $\ell$ must be provided such that $H$ neither contains a clique nor an independent set of size greater than $\ell$, but also such that each vertex of $H$ is contained in an independent set or a clique of size exactly $\ell$. The magnitude of $\ell$ in comparison to the magnitude of $t$ plays a crucial role for the quality of our construction.

We emphasize that the requirements on the host graph are loosened slightly compared to the example of Section 1. We achieve this by embedding each instance first in another local structure then to be embedded in the host graph.

Given a host graph $H$ on $t'$ vertices and graphs $G_1, \ldots, G_t$ with $t \leq t'$ we construct a graph $G' = Embed(H, k; G_1, \ldots, G_t)$, the embedding of the graphs $G_i$ into the graph $H$, as follows. We use the dummy graph $D_c$ that is defined as the join of a $(c-2)$-clique with an independent set of size $c-1$. Note that $\alpha(D_c) = \omega(D_c) = c-1$. Now, assign each instance $(G_i, k)$ to a unique vertex of $H$. By possibly repeating instances we achieve that each vertex of $H$ is assigned an instance. For each assignment of an instance $G_i$ to a vertex $v$ of $H$ create a local graph $H_v$ obtained by joining a copy of $G_i$ to a copy of $D_{k-1}$, joining a copy of the complement $\overline{G_i}$ to another copy of $D_{k-1}$, and then forming the disjoint union of the two joins. Finally, to obtain $G'$, we connect all graphs $H_v$ according to the adjacency in $H$: We fully connect $H_v$ and $H_{v'}$ if and only if $v$ and $v'$ are adjacent vertices of $H$.

The fact that we may obtain different embeddings, by assigning the instances in a different fashion to the vertices of the host graph will be irrelevant for our purposes.

**Lemma 4.** *Let $H$ be a host graph on $t'$ vertices and $(G_1, k), \ldots, (G_t, k)$ legal inputs with $t \leq t'$ for* REFINEMENT RAMSEY$(k)$. *Suppose every vertex of $H$ is contained in a clique of size $\ell$ or an independent set of size $\ell$ but $H$ neither contains a clique nor an independent set of size $\ell + 1$, then $Embed(H, k; G_1, \ldots G_t)$ has a clique or an independent set of size $\ell \cdot (2k-2)$. Furthermore, it contains a clique or an independent set of size $\ell \cdot (2k-2) + 1$ if and only if $(G_i, k)$ is a **yes** instance for some $i \in \{1, \ldots, t\}$.*

*Proof.* It is easy to see that the local structures $H_v$ from the Embedding construction contain both cliques and independent sets of size $2k-2$. Furthermore, if an instance $(G_i, k)$ is a **yes** instance, then the graph $H_v$ contains both an independent set and a clique of size $2k-1$ (in both cases using $k-1$ vertices from one of the two copies of $D_{k-1}$).

Suppose $V' \subseteq V(H)$ forms a clique of size $\ell$ in $H$. We can choose a clique of size $2k-2$ in every local graph $H_v$ that is assigned to vertex $v \in V'$. The union of these cliques forms a clique of size $\ell \cdot (2k-2)$ in $Embed(H, k; G_1, \ldots G_t)$. The analogous statement is true for independent sets.

Since every vertex in $H$ is contained in a clique or an independent set of size $\ell$, if some $(G_i, k)$ is a **yes** instance, then we can choose a clique or an independent set of size $2k - 2 + 1$ in $H_v$, where $v$ is the vertex of $H$ to which $(G_i, k)$ is assigned to, and thus in total we obtain clique or an independent set of size $\ell \cdot (2k-2) + 1$ in $Embed(H, k; G_1, \ldots G_t)$.

Finally, if no instance $(G_i, k)$ is a **yes** instance then no clique and no independent set in the graph $Embed(H, k; G_1, \ldots G_t)$ can contain more than $2k-2$ vertices from the same local graph $H_v$. Since no clique or independent set in $Embed(H, k; G_1, \ldots G_t)$ can contain vertices from more than $\ell$ different local graphs, $Embed(H, k; G_1, \ldots G_t)$ contains neither a clique nor an independent set of size $\ell \cdot (2k-2) + 1$. □

## 5 A kernelization lower bound for Ramsey($k$)

In this section we derive our kernelization lower bound for RAMSEY($k$). The main work lies in developing a co-nondeterministic composition algorithm for REFINEMENT RAMSEY($k$). Using the embedding construction of the previous section, this is centered around finding a suitable host graph. The following lemma about gaps between consecutive Ramsey numbers is required to ensure that such a graph can be found. We remark that a general result for additive or even multiplicative gaps that holds for any pair of consecutive (diagonal) Ramsey numbers is not known. All logarithms are base 2, and we take $\log t$ to be at least 1 for $t \geq 0$.

**Algorithm 1** *Compose*

**Input:** $t$ instances $(G_1, k), \ldots, (G_t, k)$ of RAMSEY$(k)$
**Output:** "**yes**" or an instance $(G', k')$ with $k' = \mathcal{O}(\log t \cdot k)$.

1: If $k < 3$ then solve each instance in time $\mathcal{O}(n^{k+2}) = \mathcal{O}(n^4)$ and answer accordingly.
2: Guess integers $T \in \{1, \ldots, (\lceil 8 \log t \rceil + 1) \cdot t\}$ and $\ell \in \{1, \ldots, \lceil 8 \log(t) \rceil\}$.
3: Guess a host graph $H$ with $T$ vertices.
4: Guess $t$ vertex sets $A_1, \ldots, A_t \in \binom{V(H)}{\ell}$, which are allowed to overlap.
5: Unless all $A_i$ induce independent sets or cliques and their union has size at least $t$, **return yes**.

6: Let $A'$ denote an arbitrary minimal choice of sets $A_i$ such that their union has size at least $t$.
7: Let $H' = H[A']$.
8: Let $G' = Embed(H', k; G_1, \ldots, G_t)$.
9: **return** $(G', k')$ where $k' := \ell \cdot (2k - 2) + 1$.

**Lemma 5.** *For every integer $t > 3$ there exists an integer $\ell \in \{1, \ldots, \lceil 8 \log(t) \rceil\}$ such that $R(\ell+1) > R(\ell) + t$.*

*Proof.* We assume the statement of the lemma is not true, then $R(\lceil 8 \log(t) \rceil + 1) \leq t \cdot \lceil 8 \log(t) \rceil + R(1)$. We use Erdős' classical bound on the Ramsey number which shows that $R(N) \geq 2^{(N-1)/2}$ for all $N \in \mathbb{N}$. This gives us $R(\lceil 8 \log(t) \rceil + 1) \geq 2^{\lceil 8 \log(t) \rceil / 2} \geq 2^{4 \log(t)} = t^4$. Assembling the two inequalities we get $t^4 \leq t \lceil 8 \log(t) \rceil + R(1)$, which is false for $t > 3$ since $R(1) = 1$. □

We now give a co-nondeterministic algorithm *Compose* (see Algorithm 1) that given $t$ instances $(G_1, k), \ldots, (G_t, k)$ of REFINEMENT RAMSEY$(k)$ will on each computation path return either the answer **yes** or a single instance $(G', k')$ with $k' = \mathcal{O}(\log t \cdot k)$. (The answer **yes** may be replaced by any constant size **yes**-instance.) We will then show that *Compose* is a co-nondeterministic composition for REFINEMENT RAMSEY$(k)$. As usual, "guessing" some integer or structure in the algorithm corresponds to a (co-)nondeterministic branching of the computation into one independent path for each possible value that the integer can take or possible structure that can occur.

**Lemma 6.** *Compose is a co-nondeterministic composition for* REFINEMENT RAMSEY$(k)$.

*Proof.* Let $t$ instances $(G_1, k), \ldots, (G_t, k)$ be given. W.l.o.g. $k \geq 3$, otherwise we can solve all instances in deterministic polynomial time and answer accordingly. Assume for now that $t > 3$.

We will first consider the case that at least one input instance is **yes**. Clearly, it suffices to check that all instances $(G', k')$ returned by the algorithm in Step 9 are **yes** too. We have $k' = \ell \cdot (2k-2) + 1$. If the host graph used for the embedding contains an independent set or a clique of size at least $\ell + 1$, then using that each local structure contains both independent sets and cliques of size $2k - 2$ we know that $G'$ contains such a set of size at least $(\ell + 1) \cdot (2k - 2) > \ell \cdot (2k - 2) + 1$; thus $(G', k')$ is **yes**. Otherwise, it follows from the cover with independent sets and cliques of size $\ell$ that $H'$ is a suitable host graph fulfilling the requirements of Lemma 4. It then follows from the lemma that $(G', k')$ is **yes**.

The other case is that all input instances are **no**. It now suffices to show that the algorithm finds a suitable host graph on at least one computation path. Lemma 4 then ensures that the output $(G', k')$ is a **no**-instance.

Let $\ell$ denote the smallest positive integer such that $R(\ell+1) > R(\ell)+t$. According to Lemma 5, we have that $\ell \leq \lceil 8 \log t \rceil$. Furthermore, by choice of $\ell$ it follows that $R(\ell) \leq (\ell-1)t + R(1) \leq \lceil 8 \log t \rceil \cdot t$. Thus for some choice of $T \in \{1, \ldots, (\lceil 8 \log t \rceil + 1) \cdot t\}$ and $\ell \in \{1, \ldots, \lceil 8 \log t \rceil\}$ guessed by *Compose* it holds that $T = R(\ell) + t < R(\ell+1)$. It follows that there exists a graph $H$ on $T$ vertices which contains neither a clique nor an independent set on $\ell + 1$ vertices. Thus in at least one computation path of the algorithm such a graph $H$ will be found. Let us consider such a computation path and the corresponding graph $H$. (If $t \leq 3$, then $R(3) = 6$ and $R(2) = 2$ guarantees that appropriate values of $T$ and $\ell$ are found.)

As $T = R(\ell) + t$ there must exist cliques and independent sets $A_i$ each of size $\ell$ which cover at least $t$ vertices of $H$; this follows from the definition of Ramsey numbers: While there are at least $R(\ell)$ uncovered vertices, the subgraph induced by the uncovered vertices must contain an independent set or clique of size $\ell$. Clearly, $t$ sets $A_1, \ldots, A_t$ can be chosen in such a way that they cover at least $t$ vertices. Hence, in one computation path such sets $A_1, \ldots, A_t$ are found in $H$.

Thus, from Step 7 we get a graph $H'$ on at least $t$ vertices which contains no independent set or clique of size $\ell + 1$ but such that each vertex is contained in a clique or independent set of size $\ell$. Hence, by Lemma 4, the graph $G' = Embed(H', k; G_1, \ldots, G_t)$ has an independent set or a clique of size at least $k' = \ell \cdot (2k - 2) + 1$ if and only if a least one graph $G_i$ contains a independent set or clique of size at least $k$. We note that $k' = \ell \cdot (2k - 2) + 1$ is bounded by $t^{o(1)}k^{\mathcal{O}(1)}$, completing the proof. □

Now, having the co-nondeterministic composition, the following theorem is an immediate consequence of this composition and Lemma 3. NP-hardness of the unparameterized version of REFINEMENT RAMSEY($k$) follows from Theorem 1 using that nontrivial instances have $k \leq n$.

**Theorem 2.** *Unless* NP $\subseteq$ coNP/poly *and the polynomial hierarchy collapses to its third level* REFINEMENT RAMSEY($k$) *admits no polynomial kernelization.*

From Lemma 1 we get the desired lower bound for RAMSEY($k$). For completeness we sketch this argument as well (see Bodlaender et al. [5] for a more general version of transferring kernelization lower bounds via NP-completeness and the implicit Karp reduction).

**Corollary 1.** RAMSEY($k$) *does not admit a polynomial kernelization unless* NP $\subseteq$ coNP/poly.

*Proof.* Let $K$ be a polynomial kernelization for RAMSEY($k$) with polynomially bounded size $h$. It is easy to see that $K$ also gives a polynomial kernelization for REFINEMENT RAMSEY($k$): Applying $K$ to any instance $(G, k)$ of REFINEMENT RAMSEY($k$) gives an equivalent instance $(G', k')$ with $|G'|, k' \leq h(k)$ of RAMSEY($k$). Applying the reduction from Lemma 1 yields an equivalent instance $(G'', k'')$ of REFINEMENT RAMSEY($k$) such that the size of this instance is polynomial in the size of $(G', k')$ and with $k'' = k' + 1$. Thus from $K$ we get also a polynomial kernelization for REFINEMENT RAMSEY($k$), implying that NP $\subseteq$ coNP/poly. □

## 6 Conclusion

We have presented a co-nondeterministic composition for the REFINEMENT RAMSEY($k$) problem, thereby showing that RAMSEY($k$) and REFINEMENT RAMSEY($k$) do not admit polynomial kernelizations unless NP $\subseteq$ coNP/poly. On a high level, the use of co-nondeterminism allowed us to essentially guess an appropriate pattern in which to combine the given instances.

In conclusion we believe that the use of co-nondeterminism in compositions may help in resolving whether other problems like, e.g., MULTIWAY CUT and DIRECTED FEEDBACK VERTEX SET admit polynomial kernels. We mention in passing that similarly to compositions, the use of co-nondeterminism may also be of use for kernelization itself. While a polynomial coNP-kernelization that crucially uses nondeterminism can hardly be seen as practical, it is of significant theoretical interest. Indeed, a polynomial coNP-kernelization can be easily seen to exclude coNP-compositions as well as weak compositions (the latter depending of course on the degree of the size bound), assuming NP $\not\subseteq$ coNP/poly; the key point is that a coNP-kernelization together with a coNP-composition gives an oracle communication protocol with co-nondeterministic first player.

## Acknowledgement

The author is grateful to Pascal Schweitzer for providing Lemma 5 and for many useful comments on the paper.